\documentclass[journal, twocolumn]{IEEEtran}

\IEEEoverridecommandlockouts

\usepackage{cite}
\usepackage{amsmath,amssymb,amsfonts}
\usepackage{gensymb}
\usepackage{float}
\usepackage{stfloats}

\usepackage{array}
\usepackage{algorithmic}
\usepackage{graphicx}
\usepackage{textcomp}
\usepackage{xcolor}
\usepackage[hidelinks]{hyperref}
\usepackage{cancel}
\usepackage{multirow}

\usepackage{amsthm}
\theoremstyle{definition}

\ifCLASSOPTIONcompsoc
    \usepackage[caption=false, font=normalsize, labelfont=sf, textfont=sf]{subfig}
\else
\usepackage[caption=false, font=footnotesize]{subfig}
\fi

\DeclareMathOperator{\sinc}{sinc}
\DeclareMathOperator{\rect}{rect}

\newtheorem{approximation}{Approximation}

\begin{document}

\title{Approximation of the Range Ambiguity Function in Near-field Sensing Systems}

\author{
    Marcin Wachowiak,~\IEEEmembership{Member,~IEEE,} 
    André Bourdoux,~\IEEEmembership{Senior~Member,~IEEE,}
    Sofie Pollin,~\IEEEmembership{Member,~IEEE,}%
    
    \thanks{
        Marcin Wachowiak and Sofie Pollin are with imec, 3001 Leuven, Belgium and also with the Katholieke Universiteit Leuven, 3000 Leuven, Belgium (e-mail: marcin.wachowiak@imec.be)}%
    \thanks{
        André Bourdoux is with imec, 3001 Leuven, Belgium. (Corresponding author: \textit{Marcin Wachowiak})}
}

\maketitle

\begin{abstract}

    This paper investigates the range ambiguity function of near-field systems where bandwidth and near-field beamfocusing jointly determine the resolution. 
    First, the general matched filter ambiguity function is derived and the near-field array factors of different antenna array geometries are introduced. 
    Next, the near-field ambiguity function is approximated as a product of the range-dependent near-field array factor and the ambiguity function due to the utilized waveform and bandwidth. An approximation criterion based on the aperture-bandwidth product is formulated, and its accuracy is examined. 
    Finally, the improvements to the ambiguity function offered by the near-field beamfocusing, as compared to the far-field case, are presented.
    The performance gains are evaluated in terms of resolution improvement offered by beamfocusing, peak-to-sidelobe and integrated-sidelobe level improvement for a few popular array geometries.
    The gains offered by the near-field regime are shown to be range-dependent and substantial only in close proximity to the array.
\end{abstract}


\section{Introduction}

\subsection{Problem Statement}
The need for high-throughput communications and high-resolution sensing drives the increase in antenna array sizes and their operating frequency \cite{gigantic_mimo, high_res_mimo_radar, wavefield_net_sensing}.
As a result, the radiative near-field (NF) region of the array is extended and it becomes feasible to consider users or targets located within it  \cite{fresnel_lower_bound, fresnel_for_aperture, gigantic_mimo}. In the near-field region, the phase difference across the array is modelled in full detail by a spherical wavefront. The nonlinear phase difference across antennas results in an array factor (AF) that is range-dependent, facilitating beamforming with a limited beamdepth \cite{primer_on_nf_bf, large_arr_beamdepth}. 
The nonlinear phase difference can be seen as an additional source of diversity, bringing novel capabilities and performance improvements.
In the conventional far-field systems, the array steering vector is only a function of angle, while in the near-field system, it is a function of both range and angle \cite{friedlander_loc_sig_nf}. 
Owing to this, the array is able to discern users or targets at the same angle but different distances solely on the phase information (the steering vector) and with limited reliance on bandwidth \cite{nf_beamfocusing_yonina, nf_radar_testbed}. 

The gains offered by near-field systems are seen as key enablers of the next generation of communication and sensing networks, especially in bandwidth-constrained scenarios \cite{6g_loc_and_sensing, sdma_ldma_nf_comms}.
When considering sensing and localization performance, the ambiguity function serves as a key performance metric. However, as the communication and sensing systems expand in size and become more distributed, the resultant ambiguity function becomes complex and cumbersome to evaluate without the use of numerical methods. The complex signal models, lack of closed-form expressions and computational burden often obscure the performance gains offered by the near-field regime.

\subsection{Relevant works}

The near-field array factor of different array geometries has been introduced in \cite{primer_on_nf_bf, large_arr_beamdepth, uca_near_field}. The closed-form near-field array factor offers an analytical method to evaluate the sensing performance of a narrowband near-field system.
In \cite{nf_res_approx}, an approximation of the narrowband near-field sensing resolution due to beamfocusing is provided for a uniform linear array (ULA).
Next, the gains offered by MIMO configuration on the narrowband near-field ambiguity function were analyzed in \cite{mw_nf_res_alpha}.
The sensing performance of a near-field system with bandwidth for point and extended targets was investigated in \cite{nf_af_pt_et_miso_simo} in terms of Cramér-Rao lower bound (CRLB). More analyses regarding the near-field or holographic positioning can be found in \cite{crlb_holo, clrb_exlamimo}. When evaluating the sensing performance, the CRLB metric is a golden standard; however, it is not always straightforward to obtain and is mainly derived for idealized scenarios with a single target \cite{mimo_radar_sig_proc}. A more practical and multiple target and also waveform dependent performance can be inferred from the ambiguity function of a system \cite{mimo_radar_afs, gen_af_of_mimo_sys}.
The positioning ambiguity function for an NF system with bandwidth and a user located in the interior of a circular array was studied in \cite{circ_cf_ambiguity_func}.
In \cite{nf_af_separation}, the near-field ambiguity function is approximated as a product of the near-field AF and waveform autocorrelation function, under a more stringent criterion than the one presented in this work.

\subsection{Contributions}
This work proposes a simple approximation of the range ambiguity function in the near-field of an array. The analysis focuses on the range dimension with the Doppler dependence intentionally omitted to highlight the key range-related characteristics. The total range ambiguity function is decomposed into a product of the near-field array factor (AF) and the ambiguity function due to the waveform and corresponding bandwidth (BW). The approximation criterion is derived for SIMO/MISO and MIMO configurations and is expressed in terms of a bandwidth-aperture product. The accuracy of the approximation is evaluated for different configurations and array geometries.
Separate treatment of NF AF and waveform ambiguity function allows for a straightforward assessment of the gains offered by the near-field region to a system with bandwidth and vice versa. When the near-field resolution achieved through beam focusing matches that provided by the bandwidth, the range-dependent array factor enhances the peak-to-sidelobe (PSL) and integrated-sidelobe (ISL) levels by at least the amount corresponding to the sidelobe levels of the near-field array factor. The sidelobe improvement is shown to diminish with increasing distance due to the degradation of the NF beamfocusing resolution. The performance gains due to near-field beamfocusing are evaluated across a few popular array geometries.
On the other hand, in bandwidth-limited systems, the bandwidth effectively improves the initially poor sidelobe performance of narrowband NF systems. To guarantee uniform PSL performance that matches the far-field levels, the minimum required bandwidth is computed.
Provided range ambiguity function approximation can be utilized to extend CLEAN algorithms \cite{clean_algo} for near-field scenarios and reduce their computational complexity.

The remainder of this paper is organized as follows. The signal model and the approximation are presented in Sec. \ref{sec:sig_mod}. Sec. \ref{sec:nf_af_per_aperture}. provides the NF array factors of different array geometries. In Sec. \ref{sec:af_separability_cond}, the approximation criterion is derived and the accuracy of the approximation is evaluated. Finally, the performance improvement metrics due to near-field AF are discussed in Sec. \ref{sec:perf_improv_metrics}.

\section{Signal model}
\label{sec:sig_mod}
Consider a set of $M$ transmit and $N$ receive antennas denoted as $\mathcal{M}$ and $\mathcal{N}$, which form a large transmit and receive aperture. The position of the antennas is given by $\mathbf{p}_m^{\mathrm{tx}} = \left[ x_m^{\mathrm{tx}}, y_m^{\mathrm{tx}}, z_m^{\mathrm{tx}} \right]^T$ and $\mathbf{p}_n^{\mathrm{rx}} = \left[ x_n^{\mathrm{rx}}, y_n^{\mathrm{rx}}, z_n^{\mathrm{rx}}\right]^T$. For simplicity of analysis, assume an odd number of antennas. The indexing of the antennas is assumed to be symmetric, with the $0$th antenna located in the weight center of the aperture.
Given an arbitrary array geometry, the distance from the $m$th antenna to some point in space $\mathbf{p} =\left[x, y, z\right]^T$ is
\begin{align}
    \label{eq:euclidean_distance}
    d_m^{\mathrm{tx}}(\mathbf{p}) &= \left|\left| \mathbf{p} - \mathbf{p}_m^{\mathrm{tx}} \right|\right|_2, \\
    d_n^{\mathrm{rx}}(\mathbf{p}) &= \left|\left| \mathbf{p} - \mathbf{p}_n^{\mathrm{rx}} \right|\right|_2,
\end{align}
where $\left|\left|\cdot \right|\right|_2$ denotes the Euclidean distance. 
A single point of interest is located at $\mathbf{p}' = \left[ x', y', z' \right]$. The distances to the point of interest from the transmit and receive antenna are $d_m^{\mathrm{tx}}(\mathbf{p}')$ and $d_n^{\mathrm{rx}}(\mathbf{p}')$. The bistatic distance is denoted as $d_{m,n}(\mathbf{p}') = d_m^{\mathrm{tx}}(\mathbf{p}') + d_n^{\mathrm{rx}}(\mathbf{p}')$.
The point of interest is located within the radiative near-field of the apertures, where the path loss variations across the aperture can be considered negligible \cite{primer_on_nf_bf} and the propagation channel is modeled as a line-of-sight. Under the assumption that the complex target response is invariant across all transmit-receive paths and the channel simplifies to a phase-only response
\begin{align}
    h_{m,n}(\mathbf{p}, f) &= e^{-j2\pi \frac{f}{c} d_{m,n}(\mathbf{p})},
\end{align}
where $f$ is the operating frequency and $c$ is the speed of light.

Each transmitter emits a waveform of bandwidth $B$ at a center frequency $f_{\mathrm{c}}$ with its baseband frequency domain representation given by $S_m(f)$ and power normalized to unity $\int_{-B/2}^{B/2}|S_m(f)|^2\, df = 1$. The transmitted waveforms are assumed orthogonal
\begin{align}
    \int_{-B/2}^{B/2} S_m(f) S_{m'}^*(f)\, df =\begin{cases}
    1,\quad m = m'\\
    0,\quad m \neq m'
    \end{cases}.
 \end{align}
Depending on the specific method used to achieve orthogonality, residual phase error may persist \cite{tdm_ortho_compensation}. In the following, these errors are assumed to be fully compensated.

The system performance is evaluated in terms of the matched filter ambiguity function, which is equivalent to the maximum likelihood estimator \cite{principles_of_modern_radar}.
Given the channel model and the transmit orthogonality, the matched filter (MF) response (ambiguity function) can be expressed as
\begin{align}
    \label{eq:amb_func_general}
    \mathcal{A}(\mathbf{p}', \mathbf{p}) &= \frac{1}{\sqrt{M N}} \sum_{m \in \mathcal{M}} \sum_{n \in \mathcal{N}} \int_{-\infty}^{+\infty} S_m(f) S_m^*(f) \\
    & \quad \times h_{m,n}(\mathbf{p}', f) h_{m,n}^*(\mathbf{p}, f) \, df.  \nonumber
\end{align}
The correlation of the observed channel with the hypothesis can be written as 
\begin{align}
    \label{eq:chan_corr}
    h_{m,n}(\mathbf{p}', f) h_{m,n}^*(\mathbf{p}, f)  
    &= e^{-j2\pi \frac{f}{c} \left( \Delta d_m^{\mathrm{tx}}(\mathbf{p}', \mathbf{p}) + \Delta d_n^{\mathrm{rx}}(\mathbf{p}', \mathbf{p})\right)},
\end{align}
where the distance difference between $\mathbf{p}'$ and $\mathbf{p}$ for $m$th and $n$th and antenna is
\begin{align}
    \label{eq:dist_diff_per_mth_ant}
    \Delta d_m^{\mathrm{tx}}(\mathbf{p}', \mathbf{p}) &= d_m^{\mathrm{tx}}(\mathbf{p}') - d_m^{\mathrm{tx}}(\mathbf{p}), \\
    \label{eq:dist_diff_per_nth_ant}
    \Delta d_n^{\mathrm{rx}}(\mathbf{p}', \mathbf{p}) &= d_n^{\mathrm{rx}}(\mathbf{p}') - d_n^{\mathrm{rx}}(\mathbf{p}).
\end{align}
The bistatic distance difference between $\mathbf{p}$ and $\mathbf{p}'$ observed by $m$th and $n$th antenna pair is 
\begin{align}
    \label{eq:bistatic_dist_diff}
    \Delta d_{m,n}(\mathbf{p}', \mathbf{p}) &= \Delta d_m^{\mathrm{tx}}(\mathbf{p}', \mathbf{p}) + \Delta d_n^{\mathrm{rx}}(\mathbf{p}', \mathbf{p}).
\end{align}
Considering the system operates at a center frequency $f_{\mathrm{c}}$, with wavelength $\lambda = c / f_{\mathrm{c}}$, the ambiguity function can be written as
\begin{align}
     \mathcal{A}(\mathbf{p}', \mathbf{p}) 
    =& \frac{1}{\sqrt{MN}} \sum_{m \in \mathcal{M}} \sum_{n \in \mathcal{N}} \int_{-B/2}^{B/2} \left|S_m(f)\right|^2 \nonumber \\
    & \quad \times e^{-j2\pi \frac{f_{\mathrm{c}} + f}{c} \left( \Delta d_m^{\mathrm{tx}}(\mathbf{p}', \mathbf{p}) + \Delta d_n^{\mathrm{rx}}(\mathbf{p}', \mathbf{p}) \right)} \, df.
\end{align}
Factoring out the carrier frequency terms yields
\begin{align}
    \label{eq:amb_fact_int}
    \mathcal{A}(\mathbf{p}', \mathbf{p}) =& \frac{1}{\sqrt{MN}} \sum_{m \in \mathcal{M}} e^{-j 2\pi \frac{f_{\mathrm{c}}}{c} \Delta d_m^{\mathrm{tx}}(\mathbf{p}', \mathbf{p})} \nonumber \\
    & \quad \times \sum_{n \in \mathcal{N}} e^{-j 2\pi \frac{f_{\mathrm{c}}}{c} \Delta d_n^{\mathrm{rx}}(\mathbf{p}', \mathbf{p})} \nonumber \\
    & \quad \times \int_{-B/2}^{B/2} |S_m(f)|^2 e^{-j2\pi \frac{f}{c} \Delta d_{m,n}(\mathbf{p}', \mathbf{p}) } \, df.
\end{align}
By definition, the inverse Fourier transform of $|S_m(f)|^2$ is equivalent to the autocorrelation of the transmitted signal in the time domain
\begin{align}
    R_m(\tau) = \int_{-B/2}^{B/2} |S_m(f)|^2 e^{-j2\pi f \tau} \, df.
\end{align}
Applying the time-delay property of the Fourier transform, the integral from \eqref{eq:amb_fact_int} can be expressed as
\begin{align}
    \label{eq:autocorr}
    &\int_{-B/2}^{B/2} |S_m(f)|^2 e^{-j2\pi \frac{f}{c} \Delta d_{m,n}(\mathbf{p}', \mathbf{p}) } \, df, \nonumber \\
    & \quad =R_m \left( - \frac{\Delta d_{m,n}(\mathbf{p}', \mathbf{p})}{c}\right).
\end{align}

Next, the distance to some point $\mathbf{p}$ can be written as a distance to the array center and a distance correction factor due to antenna displacement with regard to the array center
\begin{align}
    \label{eq:dist_w_corr}
     d_m^{\mathrm{tx}}(\mathbf{p}) &= d_{m=0}^{\mathrm{tx}}(\mathbf{p}) + \delta_m^{\mathrm{tx}}(\mathbf{p}), \\
     d_n^{\mathrm{rx}}(\mathbf{p}) &= d_{n=0}^{\mathrm{rx}}(\mathbf{p}) + \delta_n^{\mathrm{rx}}(\mathbf{p}).
\end{align}
This allows for rewriting the distance difference (\ref{eq:dist_diff_per_mth_ant}-\ref{eq:dist_diff_per_nth_ant}) as a sum of the distance difference with regard to the array center and a distance difference correction factor due to antenna displacement
\begin{align}
    \Delta d_m^{\mathrm{tx}}(\mathbf{p}', \mathbf{p}) &=  d_{m=0}^{\mathrm{tx}}(\mathbf{p}', \mathbf{p}) + \Delta\delta_m^{\mathrm{tx}}(\mathbf{p}', \mathbf{p}), \\
    \Delta d_n^{\mathrm{rx}}(\mathbf{p}', \mathbf{p}) &=  d_{n=0}^{\mathrm{rx}}(\mathbf{p}', \mathbf{p}) + \Delta\delta_n^{\mathrm{rx}}(\mathbf{p}', \mathbf{p}),
\end{align}
where 
\begin{align}
    \label{eq:dist_diff_diff_corr}
    \Delta \delta_m^{\mathrm{tx}}(\mathbf{p}', \mathbf{p}) &= \delta_m^{\mathrm{tx}}(\mathbf{p}') - \delta_m^{\mathrm{tx}}(\mathbf{p}), \\
    \Delta \delta_n^{\mathrm{rx}}(\mathbf{p}', \mathbf{p}) &= \delta_n^{\mathrm{rx}}(\mathbf{p}') - \delta_n^{\mathrm{rx}}(\mathbf{p}).
\end{align}
This factor can be physically interpreted as the distance difference between two points $\mathbf{p}$ and $\mathbf{p}'$ due to antenna displacement from the center of the array.
The bistatic distance difference correction factor is
\begin{align}
    \label{eq:bistatic_delta_delta}
    \Delta\delta_{m,n}(\mathbf{p}', \mathbf{p}) = \Delta\delta_{m}^{\mathrm{tx}}(\mathbf{p}', \mathbf{p}) + \Delta\delta_{n}^{\mathrm{rx}}(\mathbf{p}', \mathbf{p}).
\end{align}
This allows for an alternative representation of the distance difference per $m$th and $n$th antenna from \eqref{eq:bistatic_dist_diff}, expressed as the distance difference with regard to antenna array centers ($m=0$ and $n=0$) and the distance difference correction factor due to antenna displacement with regard to the array center.
\begin{align}
    \label{eq:dist_diff_mn_expanded}
    \Delta d_{m,n}(\mathbf{p}', \mathbf{p}) &= \Delta d_{0,0}(\mathbf{p}', \mathbf{p}) + \Delta\delta_{m,n}(\mathbf{p}', \mathbf{p}).
\end{align}
Assume that all transmitters use waveforms with an identical autocorrelation function. Given that each autocorrelation function offers the same resolution, this corresponds to the worst-case scenario, as averaging the autocorrelation function across different transmitters would improve the system's sidelobe performance.
By replacing the integral in \eqref{eq:amb_fact_int} with the autocorrelation function from \eqref{eq:autocorr} and then expanding its the argument with \eqref{eq:dist_diff_mn_expanded} the ambiguity function becomes
\begin{align}
    \label{eq:af_autocorr_dep}
    \mathcal{A}(\mathbf{p}', \mathbf{p}) =& \frac{1}{\sqrt{MN}} \sum_{m \in \mathcal{M}} e^{-j 2\pi \frac{f_{\mathrm{c}}}{c} \Delta d_m^{\mathrm{tx}}(\mathbf{p}', \mathbf{p})} \nonumber \\
    & \quad \times \sum_{n \in \mathcal{N}} e^{-j 2\pi \frac{f_{\mathrm{c}}}{c} \Delta d_n^{\mathrm{rx}}(\mathbf{p}', \mathbf{p})} \\
    & \quad \times R \left( - \frac{\Delta d_{0,0}(\mathbf{p}', \mathbf{p})}{c} - \frac{\Delta\delta_{m,n}(\mathbf{p}', \mathbf{p})}{c}\right). \nonumber
\end{align}
To separate the ambiguity function \eqref{eq:af_autocorr_dep} into independent components, the dependency of the autocorrelation function on the antenna indices has to be negligible.
The antenna-dependent distance difference $\frac{\Delta\delta_{m,n}(\mathbf{p}', \mathbf{p})}{c}$ introduces a horizontal shift of the autocorrelation function.
To determine the constraints under which the horizontal shift of any band-limited autocorrelation function can be considered negligible, a worst-case scenario must be considered. The worst-case scenario is obtained by specifying the autocorrelation function with the narrowest peak (mainlobe), i.e., the one providing the best resolution. The autocorrelation function that guarantees the best resolution when the signal is bandlimited is the sinc function \cite{spec_analysis_of_sig}. The worst case is obtained by transmitting a sinc pulse, which corresponds to a uniform power spectrum in the frequency domain.
In the worst-case, the transmitted signal in the frequency domain is $|S_m(f)|^2 = \frac{1}{B} \rect{\left( \frac{f}{B} \right)}.$ 
The correlation function for all $m$ transmitters is
\begin{align}
    &R^{\mathrm{wc}}(\tau) = \frac{1}{B} \int_{-B/2}^{B/2} e^{-j2\pi f \tau} \, df = \sinc{(B\tau)}
\end{align}
where $\sinc{(x)} = \sin{(\pi x)} / (\pi x)$.
Substituting the autocorrelation function arguments from \eqref{eq:af_autocorr_dep} yields
\begin{align}
    &R^{\mathrm{wc}}\left( - \frac{\Delta d_{0,0}(\mathbf{p}', \mathbf{p})}{c} - \frac{\Delta\delta_{m,n}(\mathbf{p}', \mathbf{p})}{c}\right) \\
    &\quad = \sinc{\left( - \frac{B}{c}\Delta d_{0,0}(\mathbf{p}', \mathbf{p})  - \frac{B}{c}\Delta\delta_{m,n}(\mathbf{p}', \mathbf{p})\right)}.
\end{align}

\begin{approximation}
\label{ass:neg_bw_impact}
To remove the antenna dependence in the autocorrelation function, the contribution of the term $\Delta \delta_{m,n}(\mathbf{p}', \mathbf{p})$ in the argument of the $\sinc{}$ function must be negligible
\begin{align}
    \label{eq:bw_constraint}
    \frac{B}{c} | \Delta\delta_{m,n}(\mathbf{p}', \mathbf{p})|  \ll 1.
\end{align}
\end{approximation}
This criterion will impose a constraint on the maximum bandwidth-aperture product for which the approximation holds. A detailed analysis of the constraints per array geometry can be found in Sec. \ref{sec:nf_af_per_aperture}.
Assuming the condition in \eqref{eq:bw_constraint} is satisfied the following approximation can be made 
\begin{align}
  R{\left( -\frac{ \Delta d_{m,n}(\mathbf{p}', \mathbf{p}) }{c}\right)} 
  & \overset{A1}{\approx} R{\left( - \frac{\Delta d_{0,0}(\mathbf{p}', \mathbf{p})}{c}  \right)}.
\end{align}
The constraint was derived for the worst-case scenario, corresponding to the autocorrelation function most sensitive to the horizontal shifts. Consequently it if the condition is satisfied, the approximation remains valid for any autocorrelation function.
Applying the approximation to \eqref{eq:af_autocorr_dep} allows separating the ambiguity function into independent terms
\begin{align}
    \mathcal{A} (\mathbf{p}', \mathbf{p})
    \overset{A1}{\approx} & \underbrace{
    R\left( - \frac{\Delta d_{0,0}(\mathbf{p}', \mathbf{p})}{c} \right)   
    }_{\chi_B(\mathbf{p}', \mathbf{p})} \nonumber \\
    & \times \underbrace{\frac{1}{\sqrt{M}} \sum_{m \in \mathcal{M}} e^{-j 2\pi \frac{f_{\mathrm{c}}}{c} \Delta d_m^{\mathrm{tx}}(\mathbf{p}', \mathbf{p})}}_{\mathrm{AF}_{\mathcal{M}}(\mathbf{p}', \mathbf{p})} \\
    & \times \underbrace{\frac{1}{\sqrt{N}}\sum_{n \in \mathcal{N}} e^{-j 2\pi \frac{f_{\mathrm{c}}}{c} \Delta d_n^{\mathrm{rx}}(\mathbf{p},  \mathbf{p}')}}_{\mathrm{AF}_{\mathcal{N}} (\mathbf{p}', \mathbf{p})}. \nonumber
\end{align}
The near-field ambiguity function can be approximated as a product of the autocorrelation function due to the utilized waveform and array factors of transmit and receive apertures, similarly to the far-field scenario \cite{mimo_radar}
\begin{align}
    \mathcal{A} (\mathbf{p}', \mathbf{p})
    \overset{A1}{\approx} & \chi_B(\mathbf{p}', \mathbf{p}) \mathrm{AF}_{\mathcal{M}} (\mathbf{p}', \mathbf{p}) \mathrm{AF}_{\mathcal{N}} (\mathbf{p}', \mathbf{p}).
\end{align}
The power of the ambiguity function or matched filter is
\begin{align}
    \label{eq:amb_func_pwr}
    |\mathcal{A} (\mathbf{p}', \mathbf{p})|^2 
    =& \left| \chi_B(\mathbf{p}', \mathbf{p}) \right|^2 \left| \mathrm{AF}_{\mathcal{M}} (\mathbf{p}', \mathbf{p}) \right|^2  \nonumber \\
    & \times \left| \mathrm{AF}_{\mathcal{N}} (\mathbf{p}', \mathbf{p}) \right|^2. 
\end{align}

The ambiguity function determines key performance metrics of a sensing system, such as resolution, peak-to-sidelobe and integrated sidelobe levels. 
By separating the ambiguity function into a product of independent components \eqref{eq:amb_func_pwr} the contribution of each term can be easily evaluated. 
This allows for a straightforward assessment of how the near-field range-dependent array factors can improve the ambiguity function of a system with a predefined bandwidth. Conversely, for a fixed large-aperture near-field sensing system, the approximation \eqref{eq:amb_func_pwr} enables the evaluation of how an increase in bandwidth improves the overall ambiguity function. The performance improvement metrics are discussed in detail in the Sec. \ref{sec:perf_improv_metrics}.

\section{Aperture geometries}
\label{sec:nf_af_per_aperture}
The near-field array factor provides sensing resolution through the range-dependent beamfocusing. The resolution, sidelobe shape and level are determined by the aperture size and geometry. The closed-form NF AF provides insight into the resolution and sidelobe levels that can be achieved solely by a large aperture.
Assuming the approximation in \eqref{eq:amb_func_pwr} holds, the closed-form NF AF enables easy evaluation of performance gains for systems with bandwidth.
In the following, closed-form expressions for range-dependent NF AF are derived for several standard antenna array geometries. 
Considered antenna geometries are (1D) uniform linear arrays (ULA), uniform circular arrays (UCA) and (2D) uniform planar rectangular arrays (URA) and uniform planar circular arrays (UPCA). The array factors are calculated for a set $\mathcal{M}$  of isotropic antennas with $|\mathcal{M}| = M$ elements.
For a fair comparison with regard to NF beamfocusing performance, each array is assumed to have the same aperture $D$, which is measured as the largest dimension. For a planar rectangular array, it is the diagonal; for a planar circular array, it is the diameter.
The general form of the array factor gain is
\begin{align}
    \label{eq:arr_fac_gen}
    \left| \mathrm{AF}_{\mathcal{M}}(\mathbf{p}', \mathbf{p}) \right|^2 = \frac{1}{M} \left| \sum_{m \in \mathcal{M}} e^{-j 2\pi \frac{f_{\mathrm{c}}}{c} \Delta d_m^{\mathrm{tx}}(\mathbf{p}', \mathbf{p})} \right|^2.
\end{align}

The points in space can be described either by Cartesian coordinates $\mathbf{p} = [x, y, z]$ or by their spherical equivalents $\mathbf{p} = [d, \theta, \phi]$, where $\theta$ is the elevation angle and $\phi$ is the azimuth angle. The origin of both polar and Cartesian coordinate systems corresponds to the centre of mass for the array.

\subsection{Uniform Linear Array (ULA)}
Consider the aperture is implemented as a uniform linear array. The spacing between the antenna elements is $d_{\mathrm{a}}$ and the antennas are distributed along the X axis $\mathbf{p}_m^{\mathrm{tx}} = \left[ m d_{\mathrm{a}}, 0 \right]^T$. The $z$ dimension is dropped as it is redundant in this setup.
The distance between the $m$th antenna and some point of interest expressed in polar coordinates $\mathbf{p}=[d, \phi]^T$ can be written as
\begin{align}
    \label{eq:ula_dist_full_form}
    d_m^{\mathrm{tx}}(\mathbf{p}) 
    &= d \sqrt{1 + \left(\frac{m d_{\mathrm{a}}}{d}\right)^2 - \frac{2(m d_{\mathrm{a}}) \cos{(\phi)}}{d}}.
\end{align}

\begin{approximation}
\label{ass:fr_dist_approx} 
The minimum distance to some point of interest from the antenna array is $1.2D$. This allows for utilizing the Taylor approximation with good accuracy \cite{fresnel_lower_bound, primer_on_nf_bf}.
The distance is approximated by Taylor expansion as follows
\begin{align}
    \label{eq:taylor_approx} 
    \sqrt{1+x} \overset{A2}{\approx} 1 + \frac{x}{2} - \frac{x^2}{8}.
\end{align} 
\end{approximation}
Ignoring the terms that are scaled by $1/d^2$ and $1/d^3$ due to their negligible contribution allows us to write the distance in \eqref{eq:ula_dist_full_form} as
\begin{align}
    \label{eq:ula_dist_taylor_approx}
    d_m^{\mathrm{tx}}(\mathbf{p}) 
    &\overset{A2}{\approx} d - md_{\mathrm{a}} \cos{(\phi)} + \frac{(m d_{\mathrm{a}})^2 \sin^2{(\phi)}}{2d}.
\end{align}
The system operates in the near-field region, where the array factor exhibits an additional dependence on the distance. To investigate how the array factor evolves over distance, the angles $\phi$ and $\phi'$ are set to be equal, allowing for the acquisition of a distance cross-cut of the AF.
The distance difference between $\mathbf{p}'$ and $\mathbf{p}$ located at the same angle $\phi'$ for $m$th element of ULA is 
\begin{align}
    \label{eq:ula_dist_diff}
    \Delta d_m^{\mathrm{tx}}(\mathbf{p}', \mathbf{p})
    &=d' - d + \frac{(m d_{\mathrm{a}})^2 \sin^2(\phi')}{2} \frac{d - d'}{d d'}.
\end{align}
The near-field array factor of the ULA is obtained by substituting the distance difference from \eqref{eq:ula_dist_diff} into the general array factor formula from \eqref{eq:arr_fac_gen}.

\begin{approximation}
\label{ass:af_int_approx}
For considered aperture geometries, the distance difference function and the channel in \eqref{eq:arr_fac_gen} vary smoothly w.r.t. to the position along the aperture. The antenna element spacing satisfies the Nyquist sampling criterion ($d_{\mathrm{a}} \leq 0.5 \lambda$), preventing the aliasing. If the array aperture is sufficiently large, the discrete AF sum can be accurately approximated by an integral over the aperture \cite{van_trees_arrays, proakis_dsp, nf_sum_to_int}. This is equivalent to approximating the discrete Fourier transform (DFT) with the continuous Fourier Transform (FT).
\end{approximation}
\begin{align}
    & \left| \mathrm{AF}_{\mathrm{ULA}}(\mathbf{p}', \mathbf{p}) \right|^2 
    = \nonumber \\
    & \quad  = \frac{1}{M} \left| \sum_{-(M-1)/2}^{(M-1)/2} e^{-j \frac{2\pi}{\lambda} \left( d' - d + \frac{(m d_{\mathrm{a}})^2 \sin^2(\phi')}{2} \frac{d - d'}{d d'} \right) } \right|^2  \\
    & \quad \overset{A3}{\approx} M \left| \frac{1}{M-1} \int_{-(M-1)/2}^{(M-1)/2} e^{-j \frac{\pi}{\lambda} (m d_{\mathrm{a}})^2 \sin^2(\phi') \frac{d - d'}{d d'} } \, dm \right|^2. \nonumber
\end{align}
Solving the integral, a closed-form expression is obtained
\begin{align}
    &\left| \mathrm{AF}_{\mathrm{ULA}}(d', d) \right|^2 = \\ 
    &= M \frac{4}{ d_{\mathrm{FA}} d_{\mathrm{ver}} }
    \left( 
    C^2\left( \sqrt{\frac{ d_{\mathrm{FA}} d_{\mathrm{ver}}}{4}} \right) + S^2\left( \sqrt{\frac{ d_{\mathrm{FA}} d_{\mathrm{ver}}}{4}} \right) 
    \right) \nonumber
\end{align}
$d_{\mathrm{ver}} = \left| \frac{1}{d'}- \frac{1}{d} \right| = \frac{|d - d'|}{d'd}$ is the absolute value of the vergence difference \cite{geom_optics}. The array aperture is calculated as $D = (M-1) d_{\mathrm{a}}$, $D_{\mathrm{eff}} = D \sin(\phi')$ is the effective array aperture at an angle $\phi'$, $C(x) = \int_{0}^{x} \cos{\left(\pi x^2 / 2 \right)} \, dx $ and $S(x) = \int_{0}^{x} \sin{\left(\pi x^2 / 2 \right)} \, dx$ are Fresnel integrals and $d_{\mathrm{FA}} = 2D_{\mathrm{eff}}^2 / \lambda$ is the Fraunhofer distance of the array. The factor $M$ scales the array gain according to the number of antennas.
Notice that for angles other than $\phi' = \pi/2$, there is a loss in the effective aperture, reducing the Fraunhofer distance and degrading beamfocusing capabilities.

\subsection{Uniform Circular Array (UCA)}
In a uniform circular array, the antenna elements are distributed on a circle with a diameter of $D$. The antenna locations are given by polar coordinates $\mathbf{p}_m^{\mathrm{tx}} = [R, \phi_m]^T$, where $\phi_m = 2\pi m / M$ and $R=D/2$ is the radius of the circle.
The $0$th element corresponds to the antenna on the X-axis (front). The distance to some point given in polar coordinates $\mathbf{p}=[d, \phi]^T$ from the $m$th antenna is
\begin{align}
    \label{eq:uca_dist_full_form}
    d_m^{\mathrm{tx}}(\mathbf{p}) 
    &= \sqrt{d^2 + R^2 - 2 d R \cos{(\phi - \phi_m)}}.
\end{align}
Applying the Taylor approximation from \eqref{eq:taylor_approx}, the distance can be approximated as  
\begin{align}
    d_m^{\mathrm{tx}}(\mathbf{p}) 
    & \overset{A2}{\approx} d - R \cos{(\phi - \phi_m)} + \frac{R^2 \sin^2{(\phi - \phi_m)}}{2d}.
\end{align}
Setting equal $\phi$ and $\phi'$ angles to obtain the distance cross-cut of the AF, the distance difference between $\mathbf{p}$ and $\mathbf{p}'$ for the $m$th element of UCA becomes 
\begin{align}
    \label{eq:uca_dist_diff}
    \Delta d_m^{\mathrm{tx}}(\mathbf{p}', \mathbf{p})
    &=d' - d + \frac{R^2 \sin^2(\phi' - \phi_m)}{2} \frac{d - d'}{d d'}.
\end{align}
The near-field array factor of the UCA is obtained by substituting the distance difference from \eqref{eq:uca_dist_diff} into the general array factor formula from \eqref{eq:arr_fac_gen}. Similarly to before, given that the antenna spacing satisfies the Nyquist sampling criterion ($d_{\mathrm{a}} \leq 0.5 \lambda$), the aperture can be assumed to be continuous, and the array factor sum can be expressed as an integral
\begin{align}
    &\left| \mathrm{AF}_{\mathrm{UCA}}(\mathbf{p}', \mathbf{p}) \right|^2 = \nonumber \\
    &\qquad\quad = \frac{1}{M} \left| \sum_{m=1}^{M} e^{-j \frac{2\pi}{\lambda} \left( d' - d + \frac{R^2 \sin^2(\phi' - \phi_m)}{2} \frac{d - d'}{d d'} \right) } \right|^2 \\
    &\qquad\quad \overset{A3}{\approx} M \left| \frac{1}{2\pi} \int_{0}^{2\pi} e^{-j \frac{2\pi}{\lambda} \left( \frac{R^2 \sin^2(\phi' - \phi_m)}{2} \frac{d - d'}{d d'} \right) } \, d\phi_m \right|^2. \nonumber
\end{align}
Solving the integral, a closed-form expression for the UCA AF is obtained
\begin{align}
    \left| \mathrm{AF}_{\mathrm{UCA}}(d', d) \right|^2 &= M \left| J_0 \left( \frac{\pi d_{\mathrm{FA}} d_{\mathrm{ver}}}{16} \right) \right|^2,
\end{align}
where $J_0$ denotes the zero-order Bessel function of the first kind.

\subsection{Uniform Rectangular Array (URA)}

Consider that the aperture is implemented as a uniform rectangular antenna array. The spacing between the antenna elements is $d_{\mathrm{a}}$ and the antennas are distributed on the XY plane with the antenna positions given by $\mathbf{p}_{m_x, m_y} = \left[m_x d_{\mathrm{a}}, m_y d_{\mathrm{a}}, 0 \right]^T$. 
The distance between the $m$th antenna and some point of interest expressed in polar coordinates $\mathbf{p}=[d, \theta, \phi]^T$ can be written as
\begin{align}
    \label{eq:ura_dist_full_form}
    d_{m_x, m_y}(\mathbf{p})
    &= \sqrt{
    \begin{aligned}
        & d^2 + (m_x^2 + m_y^2) d_{\mathrm{a}}^2 \\
        & - 2 m_x d d_{\mathrm{a}} \sin{(\theta)}  \cos{(\phi)}  \\
        &- 2 m_y d d_{\mathrm{a}} \sin{(\theta)}  \sin{(\phi)}
    \end{aligned}
    }.
\end{align}
Applying the Taylor expansion from \eqref{eq:taylor_approx}, the distance to some point of interest from the antenna with $[m_x, m_y]^T$th index can be approximated as
\begin{align}
    d_{m_x, m_y}(\mathbf{p}) 
    \overset{A2}{\approx}& d - d_{\mathrm{a}} \sin{(\theta')} \left( m_x \cos{(\phi')} + m_y \sin{(\phi')} \right) \nonumber \\
    &+ \frac{1}{2d} (m_x d_{\mathrm{a}})^2 \left(1 -\sin^2{(\theta'}) \cos^2{(\phi')} \right) \\
    &+ \frac{1}{2d} (m_y d_{\mathrm{a}})^2 \left( 1 - \sin^2{(\theta')} \sin^2{(\phi')} \right). \nonumber 
\end{align}
To analyze AF over distance, the angles between the point of interest $\mathbf{p}'$ and $\mathbf{p}$ are set to be equal, $\theta = \theta'$ and $\phi = \phi'$. The distance difference between the two points on the same line is
\begin{align}
    \label{eq:ura_dist_diff}
    &\Delta d_{m_x, m_y}(\mathbf{p}', \mathbf{p})
    = d' - d \nonumber \\
    & \qquad + \frac{1}{2} (m_x d_{\mathrm{a}})^2 \left((1 -\sin^2{(\theta'}) \cos^2{(\phi')} \right)  \frac{d-d'}{d d'} \\
    & \qquad + \frac{1}{2}(m_y d_{\mathrm{a}})^2 \left( 1 - \sin^2{(\theta'}) \sin^2{(\phi')} \right) \frac{d-d'}{d d'} \nonumber
\end{align}
The near-field array factor of the URA is obtained by substituting the distance difference from \eqref{eq:ura_dist_diff} into the general array factor formula from \eqref{eq:arr_fac_gen}. Similarly, as before, given that the antenna spacing satisfies the Nyquist sampling criterion ($d_{\mathrm{a}} \leq 0.5 \lambda$), the aperture can be assumed to be continuous and the array factor sum can be expressed as an integral. 
The array factor of URA is formulated in \eqref{eq:ura_af}.
\newcounter{MYtempeqncnt}
\begin{figure*}[!b]
    \vspace*{4pt}
    \hrulefill 
    \normalsize 
    \setcounter{MYtempeqncnt}{\value{equation}} 
    \setcounter{equation}{34} 
    \begin{align}
        \label{eq:ura_af}
        &\left| \mathrm{AF}_{\mathrm{URA}}(\mathbf{p}', \mathbf{p}) \right|^2 
        = \frac{1}{M} \left| \sum_{\substack{m_x = \\ -(M-1)/2}}^{(M-1)/2} 
        e^{-j \frac{\pi}{\lambda} (m_x d_{\mathrm{a}})^2 \left(1 -\sin^2{(\theta'}) \cos^2{(\phi')} \right) \frac{d-d'}{d d'}}
         \sum_{\substack{m_y = \\ -(M-1)/2}}^{(M-1)/2} e^{-j \frac{\pi}{\lambda} (m_y d_{\mathrm{a}})^2 \left( 1 - \sin^2{(\theta'}) \sin^2{(\phi')} \right)  \frac{d-d'}{d d'}
        } \right|^2  \\
        & \overset{A3}{\approx} M \left| 
        \frac{1}{(M-1)^2} \int_{-(M-1)/2}^{(M-1)/2} 
        e^{-j \frac{\pi}{\lambda} (m_x d_{\mathrm{a}})^2 \left(1 -\sin^2{(\theta'}) \cos^2{(\phi')} \right) \frac{d-d'}{d d'}} \, dm_x
        \int_{-(M-1)/2}^{(M-1)/2} e^{-j \frac{\pi}{\lambda} (m_y d_{\mathrm{a}})^2 \left( 1 - \sin^2{(\theta'}) \sin^2{(\phi')} \right)  \frac{d-d'}{d d'}} \, dm_y
        \right|^2 \nonumber  
    \end{align}
\end{figure*}
Solving the integral in a similar manner as for the ULA, yields
\begin{align}
    & \left| \mathrm{AF}_{\mathrm{URA}} (d', d) \right|^2 =  \\
    & \qquad M \frac{4}{ d_{\mathrm{FA}_x} d_{\mathrm{ver}} }
    \left( C^2\left( \sqrt{\frac{ d_{\mathrm{FA}_x} d_{\mathrm{ver}} }{4}} \right) 
    + S^2\left( \sqrt{\frac{ d_{\mathrm{FA}_x} d_{\mathrm{ver}} }{4}} \right) \right) \nonumber \\
    & \qquad  \times 
    \frac{4}{ d_{\mathrm{FA}_y} d_{\mathrm{ver}} }
    \left( C^2\left( \sqrt{\frac{ d_{\mathrm{FA}_y} d_{\mathrm{ver}} }{4}} \right) 
    + S^2\left( \sqrt{\frac{ d_{\mathrm{FA}_y} d_{\mathrm{ver}} }{4}} \right) \right), 
     \nonumber
\end{align}
where $d_{\mathrm{FA}_x} = 2D_{x\ \mathrm{eff}}^2 / \lambda$ is the Fraunhofer distance of the aperture along the X axis, $D_{x\ \mathrm{eff}} = D_x \left( \sqrt{1 - \sin^2{(\theta')} \cos^2{(\phi')}} \right)$ is the effective aperture in X axis.
And $d_{\mathrm{FA}_y} = 2D_{y\ \mathrm{eff}}^2 / \lambda$ is the Fraunhofer distance of the aperture along the Y axis, $D_{y\ \mathrm{eff}} = D_y \left( \sqrt{1 - \sin^2{(\theta')} \sin^2{(\phi')}} \right)$ is the effective aperture in Y axis.
Assuming the equal $D_x$ and $D_y$ results in a square array whose total aperture is measured as a diagonal, constrained to $D$, the width and height of the array can be expressed as $D_x = D_y = \frac{\sqrt{2}}{2} D$
Considering only the positions at the array broadside when the $\phi' = \pi/2$ and $\theta'=0$. The array factor of the URA for a square geometry and a point of interest at the array normal simplifies to
\begin{align}
    &\left| \mathrm{AF}_{\mathrm{URA}} (d', d) \right|^2 = \\
    & \quad  M \left( \frac{8}{ d_{\mathrm{FA}} d_{\mathrm{ver}}} \right)^2
    \left( C^2\left( \sqrt{\frac{ d_{\mathrm{FA}} d_{\mathrm{ver}}}{8}} \right) 
    + S^2\left( \sqrt{\frac{ d_{\mathrm{FA}} d_{\mathrm{ver}} }{8}} \right) \right)^2, \nonumber  
\end{align}
where $d_{\mathrm{FA}} = 2D_{\mathrm{eff}}^2 / \lambda$ is the Fraunhofer distance of the array, and $D$ is the diagonal of the array.

\subsection{Uniform Planar Circular Array (UPCA)}
Consider an array implemented as a uniform planar circular array constructed from concentric circles. The antenna locations are given by polar coordinates $\mathbf{p}_{q, m_q} = [r_q, \pi/2, \phi_{m_q}]^T$, where $r_q$ is the radius of the $q$th circle and $\phi_{m_q} = 2\pi m_q / M_q$ is the angle of the $m_q$th antenna on the ring, with $M_q$ total elements on the $q$th ring.
The distance to some point given in polar coordinates $\mathbf{p}=[d, \theta, \phi]^T$ from the $m$th antenna on the $q$th ring is
\begin{align}
    \label{eq:upca_dist_full_form}
    d_{q, m_q}(\mathbf{p}) 
    &= \sqrt{d^2 + r_q^2 - 2 d r_q \sin{(\theta)} \cos{(\phi - \phi_{m_q})}}.
\end{align}
Applying the Taylor expansion from \eqref{eq:taylor_approx}, the distance can be approximated as  
\begin{align}
    d_{q, m_q}(\mathbf{p}) 
    \overset{A2}{\approx}& d - r_q \sin{(\theta)}\cos{(\phi - \phi_{m_q})} \nonumber \\
    &+ \frac{r_q^2 \left( 1 - \sin{(\theta)} \cos^2{(\phi - \phi_{m_q})} \right)}{2d}.
\end{align}
To analyze AF over distance, the angles between the points of interest $\mathbf{p}'$ and $\mathbf{p}$ are set to be equal, $\theta = \theta'$ and $\phi = \phi'$. The distance difference between the two points of interest positioned on the same line originating from the center is
\begin{align}
    \label{eq:upca_dist_diff}
    &\Delta d_{q, m_q}(\mathbf{p}', \mathbf{p}) =  d' - d  \\
    & \qquad\qquad + \frac{1}{2} r_q^2 \left( 1 - \sin{(\theta')} \cos^2{(\phi' - \phi_{m_q})} \right) \frac{d-d'}{d d'}. \nonumber
\end{align}
The near-field array factor of the UPCA is obtained by substituting the distance difference from \eqref{eq:upca_dist_diff} into the general array factor formula from \eqref{eq:arr_fac_gen}. Given that the antenna spacing satisfies the Nyquist sampling criterion ($d_{\mathrm{a}} \leq 0.5 \lambda$), the aperture can be assumed to be continuous and the array factor sum can be expressed as an integral 
\begin{align}
    &\left| \mathrm{AF}_{\mathrm{UPCA}}(\mathbf{p}', \mathbf{p}) \right|^2 = \nonumber \\
    &\qquad = \frac{1}{M} \Bigg| \sum_{q \in \mathcal{Q}}
    \sum_{m_q\in \mathcal{M}_q}^{} 
    e^{-j \frac{2\pi}{\lambda} \left(
    d' - d + \frac{1}{2} r_q^2 \frac{d-d'}{d d'} \right) } 
    \\
    & \qquad\qquad\qquad\qquad\qquad \times e^{j \frac{2\pi}{\lambda} \left(\sin{(\theta')} \cos^2{(\phi' - \phi_{m_q})} \right) \frac{d-d'}{d d'}}
    \Bigg|^2 \nonumber \\
    & \qquad \overset{A3}{\approx} M \Bigg| \frac{1}{\pi R^2} \int_{0}^{2\pi} \int_{0}^{R} 
    r_q e^{-j \frac{2\pi}{\lambda} \left(
    d' - d + \frac{1}{2} r_q^2 \frac{d-d'}{d d'} \right) } \nonumber \\
    & \qquad\qquad\qquad \times e^{j \frac{2\pi}{\lambda} \left(\sin{(\theta')} \cos^2{(\phi' - \phi_{m_q})} \right) \frac{d-d'}{d d'}}
    \, dr_q \, d\phi_m  \Bigg|^2. \nonumber
\end{align}
For points of interest located at the array normal, $\theta = 0$, the AF integral simplifies, allowing for obtaining a closed-form expression as follows
\begin{align}
    & \left| \mathrm{AF}_{\mathrm{UPCA}}(d', d) \right|^2 = \nonumber \\
    & \qquad\quad = M \left| \frac{1}{\pi R^2} \int_{0}^{2\pi} \int_{0}^{R} 
    r_q e^{-j \frac{\pi}{\lambda}  r_q^2 \frac{d-d'}{d d'} }
    \, dr_q\, d\phi_{m_q} \right|^2  \\
    & \qquad\quad = M \sinc^2{\left( \frac{d_{\mathrm{FA}} d_{\mathrm{ver}} }{16} \right)} \nonumber
\end{align}

\begin{figure}[t]
    \centering
    \includegraphics[width=\linewidth]{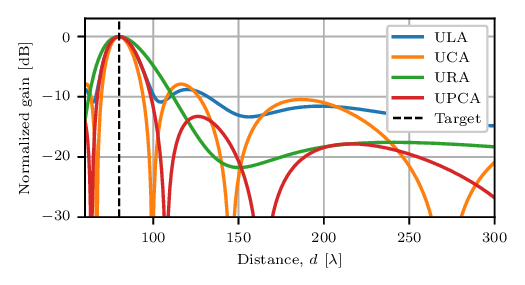}
    \caption{Array factor per geometry for $D=50 \lambda$ a target located at at $80 \lambda$.}
    \label{fig:af_per_geom}
\end{figure}

\begin{figure}[t]
    \centering
    \includegraphics[width=\linewidth]{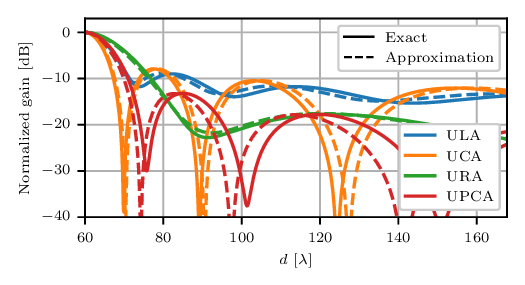}
    \caption{Comparison of the exact and approximated array factor per geometry for $D=50\lambda$ and target located at $d'_{\min} = 60\lambda$.}
    \label{fig:nf_af_exact_vs_approx}
\end{figure}

Fig. \ref{fig:af_per_geom} illustrates the array factor for each geometry. The array factor defined by the geometry clearly determines the shape of the beampattern, the 3 dB radial resolution, the sidelobe level and the maximum extent of the near-field region. \cite{mw_nf_res_alpha}. 
Note that the distance approximation \eqref{eq:taylor_approx}, which allows for obtaining a closed-form array factor in the NF, introduces errors at small distances from the array. This is due to the quadratic approximation of a spherical wavefront. The worst-case error is observed for targets at $d_{\min}$. Fig. \ref{fig:nf_af_exact_vs_approx} illustrates the difference between the exact and approximated array factors. The mainlobe of the array factor is well approximated; however, there is a noticeable shift in the null positions. The approximation error decreases as the distance from the array increases. 

\subsection{Beamforming resolution (beamdepth)}

\begin{figure}[t]
    \centering
    \includegraphics[width=\linewidth]{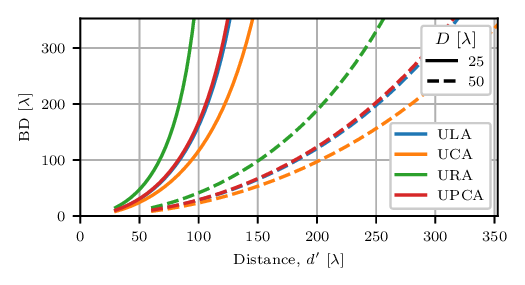}
    \caption{Beamdepth as a function of distance for two array aperture sizes and selected array geometries.}
    \label{fig:bd_vs_d}
\end{figure}

Given the closed-form AF, it is possible to calculate the 3 dB radial resolution, referred to as the beamdepth (BD).
It is obtained by computing $\alpha$, which is the value of the argument $d_{\mathrm{FA}} d_{\mathrm{ver}}$ for which the normalized AF reaches half of the maximum value \cite{mw_nf_res_alpha}. Next, by expanding the $d_{\mathrm{ver}}$ and solving for $d$, two points are the distances at which the $-$3 dB value is reached. By calculating the distance between those points, the 3 dB beamdepth $\mathrm{BD}(D, d')$ is obtained, as a function of aperture $D$ and distance $d'$; for a detailed derivation, refer to \cite{mw_nf_res_alpha}
\begin{align}
    \label{eq:3dB_bd_vs_d}
    \mathrm{BD}(D, d') &= \frac{2\alpha d_{\mathrm{FA}} d'^2}{d_{\mathrm{FA}}^2 - \alpha^2 d'^2}.
\end{align}
The values of parameter $\alpha$ per array geometry and processing are taken from \cite{mw_nf_res_alpha} and reported in Table \ref{tab:alpha_bd_min_per_geom}. MIMO/SIMO refers to the single-aperture scenario, and MIMO refers to two identical transmit and receive apertures. In the following, when referring to the MIMO system, the transmit and receive apertures are always considered identical and collocated. As the MIMO ambiguity function includes the squaring of the NF AF, the sensing beamdepth is narrower. The resolution improvement offered by MIMO can be approximated as a factor $\sqrt{2}$ \cite{mw_nf_res_alpha}.
Fig. \ref{fig:bd_vs_d} illustrates the beamdepth as a function of distance for a SIMO/MISO system. The beamdepth function increases rapidly with the distance from the array.

In the context of near-field sensing and communications, the radiative near-field is defined as a region where the array gain variations across the aperture can be considered negligible and the beamdepth is finite.  The lower limit of the near-field region is the distance at which the gain can be considered the same for all antennas across the aperture, which is $1.2D$ \cite{fresnel_lower_bound, primer_on_nf_bf}. The upper limit of the near-field region is defined as the distance where the beamdepth becomes infinite, which is $d_{\mathrm{FA}} / \alpha$, resulting in the near field region given by $[1.2D, d_{\mathrm{FA}}/\alpha]$.

The near-field beamforming resolution (BD) is primarily determined by the range $d'$ and the aperture size $D$.
The beamdepth is a monotonically increasing function, therefore, the minimum beanmdepth is obtained at a minimum distance from the array.
Consider some lower limit of the near-field distance $d' \geq 1.2D$. The beamdepth corresponding to the minimum distance from the array is observed to reach an asymptotic limit as the array aperture increases \cite{mw_nf_arr_sizing}. 
\begin{align}
    \label{eq:asymptotic_3dB_bd}
    \mathrm{BD}_{\min} = \lim_{D \to \infty} \mathrm{BD(D, d'_{\min})} = 1.44 \alpha \lambda,
\end{align}
where $d'_{\min} = 1.2 D$.
The minimum asymptotic beamdepth for $ d' = 1.2D$ and $D \to \infty$ per array geometry and processing can be found in Tab. \ref{tab:alpha_bd_min_per_geom}.

\begin{table}[htb]
    \caption{Parameter $\alpha$ and minimum $BD$ per array geometry and processing.}
    \label{tab:alpha_bd_min_per_geom}
    \centering
    \def\arraystretch{1.3}
    \begin{tabular}{
      >{\centering\arraybackslash}m{0.15\linewidth}<{}
      |>{\centering\arraybackslash}m{0.15\linewidth}<{}
      |>{\centering\arraybackslash}m{0.15\linewidth}<{}
    |>{\centering\arraybackslash}m{0.15\linewidth}<{}
      |>{\centering\arraybackslash}m{0.15\linewidth}<{}
    }
         \multirow{2}{*}{\shortstack{Array\\geometry}} & \multicolumn{2}{c|}{$\mathrm{SIMO/MISO}$} &  \multicolumn{2}{c}{$\mathrm{MIMO}$} \\
         \cline{2-5}
         & $\alpha$ & $\mathrm{BD_{\min}}\ [\lambda]$ & $\alpha$ & $\mathrm{BD_{\min}}\ [\lambda]$ \\
         \hline
         ULA & 6.952 & 10.01 & 4.969 & 7.15 \\
         \hline
         UCA & 5.737 & 8.26 & 4.148 & 5.98 \\
         \hline
         URA & 9.937 & 14.31 & 7.068 & 10.18 \\
         \hline
         UPCA & 7.087 & 10.21 & 5.103 & 7.35 \\
    \end{tabular}
\end{table}

\section{Ambiguity function separability conditions}
\label{sec:af_separability_cond}
Given the considered aperture geometries, it is crucial to revisit the condition in \eqref{eq:bw_constraint}, needed to separate the antenna array factor and waveform autocorrelation function and to determine for which parameter ranges the separation approximation holds. 
Without loss of generality, the following analysis is done with regard to the index $m$.
To define the constraint, the maximum value of the bistatic distance correction factor $\Delta\delta_m^{\mathrm{tx}}(\mathbf{p}', \mathbf{p})$ is of interest. From \eqref{eq:dist_diff_diff_corr} it is apparent that
$\Delta\delta_m^{\mathrm{tx}}(\mathbf{p}', \mathbf{p})$ is maximized when the difference between $\mathbf{p}'$ and $\mathbf{p}$ is the greatest.
The lower limit $d_{\min}$ of the considered near-field region is $1.2D$ from Assumption \ref{ass:fr_dist_approx}. To consider a worst-case scenario and simplify the analysis, the upper distance limit $d_{\max}$ is unconstrained ($d_{\max} \to \infty$), resulting in the greatest distance difference between the center and edge antenna elements.

Substituting 
$\mathbf{p}' = \mathbf{p}_{\max} = [d_{\max}, 0,0]^T$ and $\mathbf{p} = \mathbf{p}_{\min} = [d_{\min}, 0,0]^T$ and given that the zeroth element of the antenna array coincides with the origin, $\delta_m^{\mathrm{tx}}(\mathbf{p})$ and $\delta_m^{\mathrm{tx}}(\mathbf{p}')$ from \eqref{eq:dist_w_corr} become
\begin{align}
    \delta_m^{\mathrm{tx}}(\mathbf{p}_{\min}) &= d_m^{\mathrm{tx}}(\mathbf{p}_{\min}) - d_{\min} \\
    \delta_m^{\mathrm{tx}}(\mathbf{p}_{\max}) &= d_m^{\mathrm{tx}}(\mathbf{p}_{\max}) - d_{\max}.
\end{align}
Under these limits, the $\Delta\delta_m^{\mathrm{tx}}(\mathbf{p}_{\min}, \mathbf{p}_{\max})$ reduces to the difference in distance given by the geometrical (near-field) channel model and the far-field approximation. For the planar arrays, the largest difference in distance between the two propagation models occurs at the array boresight, which allows for specifying a worst-case scenario. Considering targets at other angles (rotated arrays) would relax the constraint, resulting in reduced distance difference.

\subsubsection{ULA, URA and UPCA}

For a point of interest on the array boresight, the antenna elements are distributed on a plane perpendicular to it. For large $d_{\max}$, the distance difference between the $m$th antenna and the array center becomes negligible, arriving at the (planar wave) far-field approximation
\begin{align}
    \lim_{d \to \infty} \delta_m^{\mathrm{tx}}(\mathbf{p}_{\max}) =  0
\end{align}
This represents a worst-case value, as considering a smaller upper limit would reduce the error distance.
The distance difference correction term $\Delta\delta_m^{\mathrm{tx}}(\mathbf{p}', \mathbf{p})$ reduces to
\begin{align}
    \label{eq:ula_d_min_d_max}
    \Delta\delta_m^{\mathrm{tx}}(\mathbf{p}_{\min}, \mathbf{p}_{\max}) = d_m^{\mathrm{tx}}(\mathbf{p}_{\min}) - d_{\min}.
\end{align}
The value of the distance difference with regard to the array center \eqref{eq:ula_d_min_d_max} is maximized for $m$, which has the greatest displacement from the center of the array. Given that antenna elements are distributed on a plane perpendicular to the array normal for the listed aperture architectures, the maximum displacement is $0.5D$. The distance $d_m^{\mathrm{tx}}(\mathbf{p}_{\min})$ can be calculated as
\begin{align}
    \label{eq:delta_d_min_ula}
    \max{ \left\{ d_m^{\mathrm{tx}}(\mathbf{p}_{\min}) \right\} } &= \sqrt{d_{\min}^2 + (0.5D)^2} = 1.3 D.
\end{align}
Substituting \eqref{eq:delta_d_min_ula} in \eqref{eq:ula_d_min_d_max}, the worst-case error for ULA, URA and UPCA is 
\begin{align}
    \max{ \left\{ \delta_m^{\mathrm{tx}}(\mathbf{p}', \mathbf{p}) \right\} } = 1.3D - 1.2D = 0.1D.
\end{align}

\subsubsection{UCA} 
The distance difference is measured with regard to the $0$th antenna element, which for the UCA corresponds to the antenna on the X-axis (front). For large $d_{\max}$, the distance difference between the $m$th antenna and the $0$th antenna is
\begin{align}
     \lim_{d_{\max} \to \infty} \delta_m^{\mathrm{tx}}(\mathbf{p_{\max}}) 
     &= R (1 - \cos{(\phi_m))}.
\end{align}
Notice that the distance difference between the antennas still exists even under the far-field assumption, as the antenna elements are distributed on a plane parallel to the considered propagation direction.
For $d_{\min}$, the distance difference can be calculated using the law of cosines as
\begin{align}
    \delta_m^{\mathrm{tx}}(\mathbf{p_{\min}}) 
    &=R \left( \sqrt{6.76 - 4.8 \cos{(\phi_m)} } - 1.4 \right).
\end{align}
The total correction term for $\mathbf{p}_{\min}$ and $\mathbf{p}_{\max}$ is
\begin{align}
    \Delta\delta_m^{\mathrm{tx}}(\mathbf{p}_{\min},\mathbf{p}_{\max}) 
    = R \Big( \sqrt{6.76 - 4.8 \cos{(\phi_m)} } & \\
     - 2.4 + \cos{(\phi_m)} &\Big) \nonumber.
\end{align}
To obtain a maximum, a derivative of the function with respect to the angle is calculated and critical points are extracted.
\begin{align}
    \arg \underset{\phi_m}{\max}{ \left(| \Delta\delta_m^{\mathrm{tx}}(\mathbf{p}_{\min},\mathbf{p}_{\max}) |\right)} &= \cos^{-1}{\left( \frac{5}{24} \right)}.
\end{align}
Given the critical points, the maximum value of the correction term for UCA is
\begin{align}
    \max{\left\{ |\Delta \delta_m^{\mathrm{tx}}(\mathbf{p},\mathbf{p}') |\right\} } 
    &= 0.104 D
\end{align}
which is in the same order as for planar array geometries.
The preceding analysis was done in regard to a single aperture (SIMO/MISO). When considering identical transmit and receive apertures, the maximum distance difference is doubled 
\begin{align}
    \label{eq:mimo_error}
    \max{\left\{ | \Delta\delta_{m,n}(\mathbf{p}', \mathbf{p}) | \right\} } 
    = 2  \max{\left\{ | \Delta\delta_{m}^{\mathrm{tx}}(\mathbf{p}', \mathbf{p}) | \right\} }
\end{align}

\begin{figure*}[b]
    \centering
    \includegraphics[width=\linewidth]{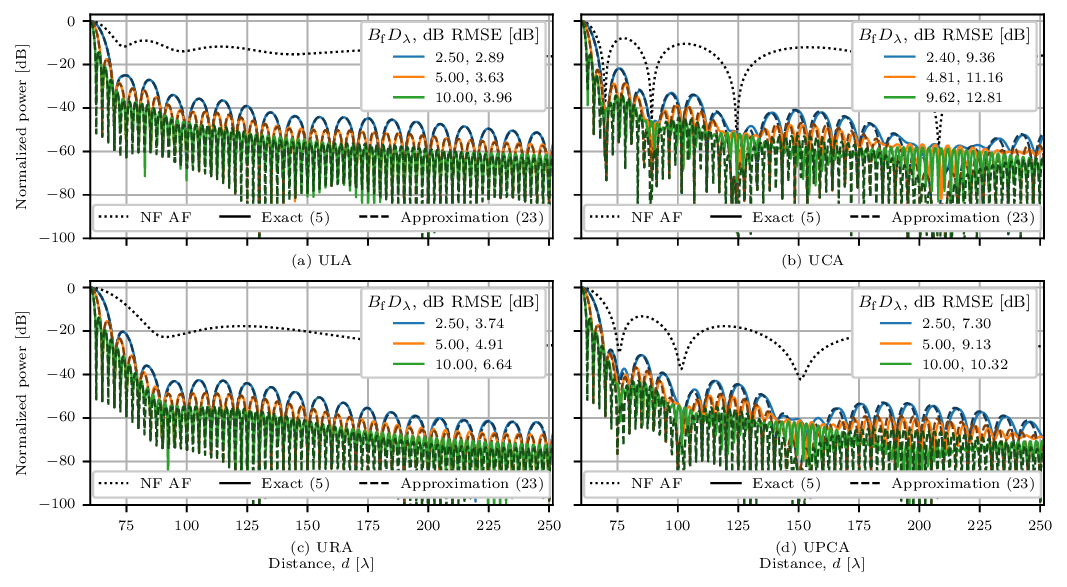}
    \caption{Comparison of the SIMO/MISO OFDM ambiguity function for different array geometries and values of $B_{\mathrm{f}} D_{\lambda}$ product, for $K=1024$, $D_{\lambda} = 50 \lambda$ and target at $d_{\min} = 60 \lambda$.}
    \label{fig:mf_vs_apx_simo_miso_combined}
\end{figure*}
\begin{figure*}[t]
    \centering
    \includegraphics[width=\linewidth]{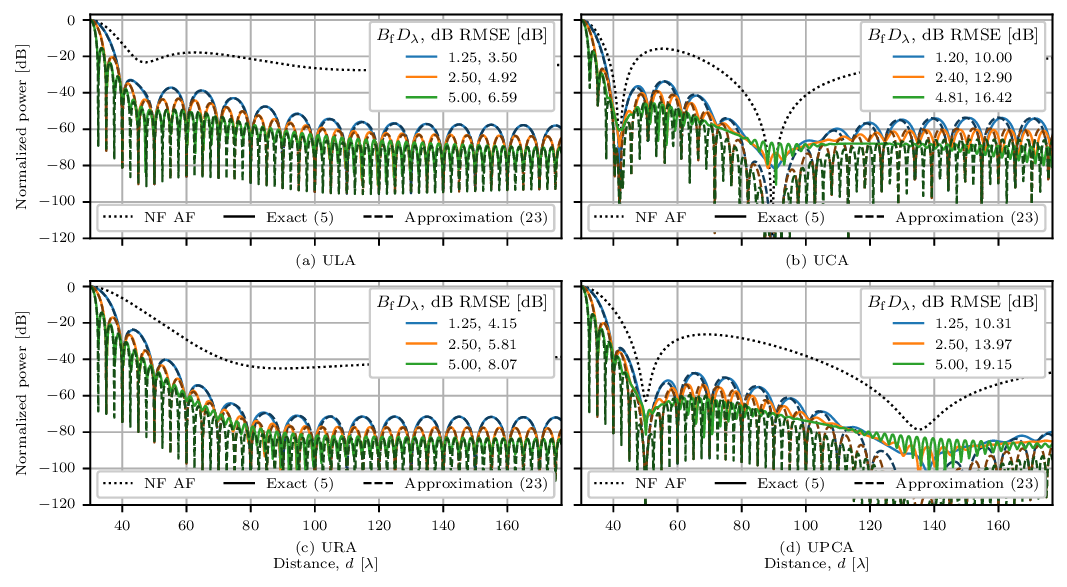}
    \caption{Comparison of the MIMO OFDM ambiguity function for different array geometries and values of $B_{\mathrm{f}} D_{\lambda}$ product, for $K=1024$, $D_{\lambda} = 25 \lambda$ and target at $d_{\min} = 30 \lambda$.}
    \label{fig:mf_vs_apx_mimo_combined}
\end{figure*}

Given the maximum, worst-case value of $\Delta\delta_{m,n}(\mathbf{p}', \mathbf{p})$ the maximum bandwidth for which the approximation holds can be calculated from \eqref{eq:bw_constraint}. 
Consider some bandwidth $B$ expressed in terms of fractional bandwidth $B = B_{\mathrm{f}}f_{\mathrm{c}}$ and aperture size expressed as a multiple of wavelength $D = D_{\lambda} \lambda$. By plugging in those expressions to \eqref{eq:bw_constraint}, the constraint on the bandwidth-aperture product is obtained
\begin{align}
    \label{eq:bfrac_dap_constraint}
    B_{\mathrm{f}}D_{\lambda} \ll \frac{D}{\max{\left\{ | \Delta\delta_{m,n}(\mathbf{p}', \mathbf{p}) | \right\} } }.
\end{align}
The bandwidth-aperture constraint specifies the range of values for which the ambiguity function can be approximated as a product of the NF AF and the ambiguity function due to the waveform.
Note that the proposed approximation is applicable to a broad range of the array factors, including sparse arrays as long as the elements are distributed on a line, plane or circle, for which the bandwidth-aperture constraints were derived.

Table \ref{tab:fbw_dap_prod_constraint} presents the bandwidth-aperture product constraints for the considered aperture geometries and processing schemes. In the MIMO system, the bandwidth-aperture product is reduced twofold due to the doubling of the maximum distance difference in a bistatic scenario \eqref{eq:mimo_error}. 
\begin{table}[htb]
    \caption{Bandwidth-aperture product constraint for different array geometries and processing.}
    \label{tab:fbw_dap_prod_constraint}
    \centering
    \def\arraystretch{1.2}
    \begin{tabular}{
      >{\centering\arraybackslash}m{0.3\linewidth}<{}
      |>{\centering\arraybackslash}m{0.25\linewidth}<{}
      |>{\centering\arraybackslash}m{0.25\linewidth}<{}
    }
         Array geometry & SIMO/MISO & MIMO \\
         \hline
         ULA, URA, UPCA & $B_{\mathrm{f}} D_{\lambda}\ll 10$ &  $B_{\mathrm{f}}D_{\lambda} \ll 5$\\
         \hline
         UCA & $B_{\mathrm{f}}D_{\lambda} \ll 9.615$ & $B_{\mathrm{f}} D_{\lambda}\ll 4.808$\\
    \end{tabular}
\end{table}
Table \ref{tab:sys_example} illustrates the resulting bandwidth-aperture product for several existing and envisoned cellular and automotive sensing systems \cite{gigantic_mimo}. Half-wavelength $(\lambda/2)$ spacing is assumed and $M$ corresponds to the number of antennas in ULA. Obtained $B_{\mathrm{f}} D_{\lambda}$ values indicate that the proposed approximation of the ambiguity function can be reliably applied to a wide range of system configurations.
\begin{table}[htb]
    \caption{Typical parameters of selected communications and automotive sensing systems.}
    \label{tab:sys_example}
    \centering
    \def\arraystretch{1.2}
    \begin{tabular}{
      >{\centering\arraybackslash}m{0.2\linewidth}<{}
      |>{\centering\arraybackslash}m{0.2\linewidth}<{}
      |>{\centering\arraybackslash}m{0.1\linewidth}<{}
      |>{\centering\arraybackslash}m{0.1\linewidth}<{}
      |>{\centering\arraybackslash}m{0.15\linewidth}<{}
    }
        $f_c$ & $B$ & $M$ & $D_{\lambda}$  & $B_{\mathrm{f}} D_{\lambda}$ \\
        \hline
        3.5 GHz & 100 MHz & 128 & 63.5 & 1.81 \\
        \hline
        7.8 GHz & 200 MHz & 256 & 127.5  & 3.27 \\
        \hline
        15 GHz & 400 MHz & 512 & 255.5 & 6.81 \\
        \hline
        28 GHz & 1 GHz & 256 & 127.5 & 4.55 \\
        \hline
        79 GHz & 4 GHz & 256 & 127.5 & 6.46 \\
    \end{tabular}
\end{table}

\subsection{OFDM}
Consider OFDM as a method to implement a near rectangular spectrum. Given an OFDM symbol consisting of $K$ subcarriers, the ambiguity function of the waveform is given by 
\begin{align}
    \chi_{B} &=\frac{\sinc{\left( \frac{2B}{c} (d-d') \right)}}{\sinc{\left( \frac{2B}{Kc} (d-d') \right)}}.
\end{align}
Given that $\pi \frac {B}{Kc} |d-d'| \ll 1$, the denominator can be considered negligible and the ambiguity function approximates that of a $\sinc{}$ pulse.

When comparing the ambiguity function approximation, obtained by separating the NF AF and the waveform ambiguity function, to the ideal matched filter, two factors affect the accuracy of the approximation. The first aspect is the accuracy of the analytical AF approximation, and the second is the bandwidth-aperture product. Here, only the impact of the bandwidth-aperture product is considered. The error introduced by the AF approximation is most noticeable for small distances from the array due to the quadratic approximation of the spherical distance difference.
Fig. \ref{fig:mf_vs_apx_simo_miso_combined} illustrates the actual MF ambiguity function and the approximation for MISO/SIMO and Fig.\ref{fig:mf_vs_apx_mimo_combined} for MIMO. In the comparison, the array aperture is kept constant while the bandwidth is increased. A reasonable accuracy for the worst case scenario is observed for $B_{\mathrm{f}} D_{\lambda}$ being a quarter of the constraint. 

\begin{figure}[t]
    \centering
    \includegraphics[width=\linewidth]{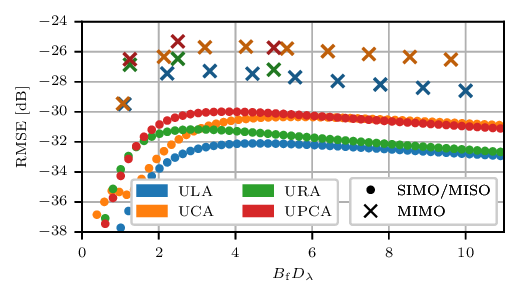}
    \caption{Linear RMSE as a function of the bandwidth-aperture product, for $K=1024$, $D_{\lambda_\mathrm{SIMO/MISO}} = 50 \lambda$, $D_{\lambda_\mathrm{MIMO}} = 25 \lambda$ and target at $d_{\min} = 1.2 D_{\lambda}$.}
    \label{fig:rmse_db}
\end{figure}
\begin{figure}[t]
    \centering
    \includegraphics[width=\linewidth]{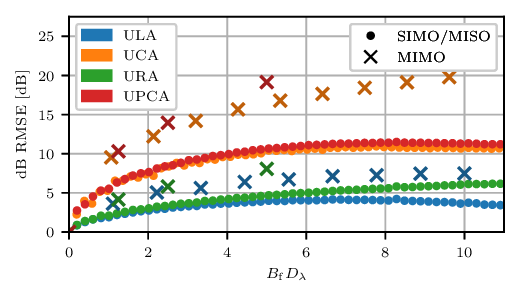}
    \caption{Decibel RMSE as a function of the bandwidth-aperture product, for $K=1024$, $D_{\lambda_\mathrm{SIMO/MISO}} = 50 \lambda$, $D_{\lambda_\mathrm{MIMO}} = 25 \lambda$ and target at $d_{\min} = 1.2 D_{\lambda}$.}
    \label{fig:db_rmse}
\end{figure}

The difference between the reference MF calculated numerically and the approximation is quantified by root mean square error (RMSE), defined as $\sqrt{\frac{1}{N} \sum_{N} \left( |\mathcal{A}_{\mathrm{MF}}|^2 - |\mathcal{A}_{\mathrm{APX}}|^2 \right)^2}$. The conventional linear RMSE captures well the differences in the high-power regions (mainlobe). To show the difference in the low-power region (sidelobes) the dB RMSE is used, which is RMSE calculated using decibel values directly.
Fig. \ref{fig:rmse_db} illustrates the linear RMSE vs bandwidth-aperture product. The low values of the linear RMSE indicate an insignificant difference between the MF and approximation in the mainlobe region. This can be attributed to the mainlobe (resolution) being quickly dominated by the bandwidth ambiguity function as the bandwidth-aperture product increases.
Fig. \ref{fig:db_rmse} shows the dB RMSE vs bandwidth-aperture product for different array geometries.
As the $B_{\mathrm{f}}D_{\lambda}$ product increases, the sidelobe levels, especially near the nulls, are also increased. Moreover, different array architectures exhibit different sensitivities to the $B_{\mathrm{f}}D_{\lambda}$. 
The impact of increased sidelobes is more apparent for architectures where the AF exhibits nulls (UCA and UPCA). 
The increase in sidelobe level is less noticeable for linear architectures such as ULA and URA.

\subsection{Minimum near-field resolution vs resolution due to bandwidth}
The considered system offers two sources of range resolution: first, due to the range-dependent array factor (beam focusing), and second, due to bandwidth. To quantify and characterize how near-field beamfocusing improves the sensing performance of a system with bandwidth and vice versa, the two resolutions must be comparable.

\begin{figure}[tb]
    \centering
    \includegraphics[width=\linewidth]{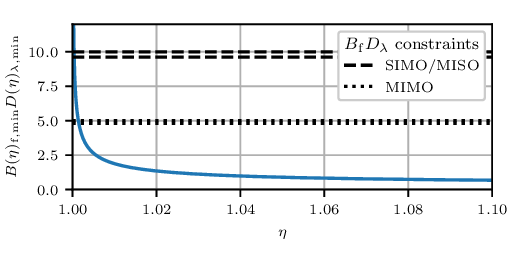}
    \caption{Fractional bandwidth, aperture product as a function of $\eta$ for equal resolution provided by beamfocusing and bandwidth.}
    \label{fig:bf_dap_prod_vs_eta}
\end{figure}

When implementing a near-field array, one may be interested in the minimum array size that guarantees a fraction of the asymptotic NF resolution, while keeping the antenna array size reasonable.
The minimum array aperture that guarantees $\eta > 1$ fraction of the asymptotic resolution \eqref{eq:asymptotic_3dB_bd} can be obtained by equating  \eqref{eq:3dB_bd_vs_d} with a fraction of \eqref{eq:asymptotic_3dB_bd} and solving for the array aperture \cite{mw_nf_arr_sizing}
\begin{align}
    \label{eq:min_d_ap_size}
    D(\eta)_{\min} = 0.6 \lambda \alpha \sqrt{\frac{\eta}{\eta - 1}}.
\end{align}
Given the aperture size and the minimum beamdepth offered by beamfocusing, let us calculate the fractional bandwidth required to provide the same resolution. To match the definitions, the range resolution is taken as the 3 dB beamwidth of the ambiguity function due to the waveform. The 3dB sensing resolution offered by a system with bandwidth $B$ is $\Delta r = \frac{0.886c}{2B}$.  Consider the $ \eta \mathrm{BD}_{\min}$ as the desired resolution, the minimum fractional bandwidth that provides the same resolution as beamfocusing is
\begin{align}
    \label{eq:bf_min}
    B(\eta)_{\mathrm{f, \min}} = \eta \frac{0.443 \lambda }{\mathrm{BD}_{\min}} 
    = \eta \frac{0.308}{\alpha}.
\end{align}
Given the size of the array and the fractional bandwidth the product of the two can be expressed as a function of $\eta$
\begin{align}
    B(\eta)_{\mathrm{f}, \min}D(\eta)_{\lambda, \min}
    &= 0.185 \eta\sqrt{\frac{\eta}{\eta - 1}}.
\end{align}
Fig. \ref{fig:bf_dap_prod_vs_eta} presents the bandwidth-aperture product of a system that guarantees a fraction of the asymptotic NF resolution $\eta$.
For $\eta = 1.01$ of the asymptotic resolution $B_{\mathrm{f}}D_{\lambda} = 1.88$ and $\eta = 1.1$ the $B_{\mathrm{f}}D_{\lambda} = 0.67$ fitting well under the constraint provided in Tab. \ref{tab:fbw_dap_prod_constraint}, making the analysis applicable to systems approaching the limits of the NF sensing resolution.

Table \ref{tab:dap_and_bfrac_for_eta} presents the minimum aperture and fractional bandwidth $B_{\mathrm{f, \min}}$ \eqref{eq:bf_min} that guarantees $\eta = 1.01$ of the asymptotic NF resolution. 
As the MIMO NF ambiguity function offers better resolution due to squaring of the NF AF, it requires a lower aperture for the same level of asymptotic resolution. Consequently, the fractional bandwidth is increased to match the MIMO resolution, keeping the product of the two constant.

\begin{table}[tb]
    \caption{Minimum aperture size $D_{\lambda, \min}$ and fractional bandwidth $B_{\mathrm{f}, \min}$ for $\eta = 1.01$,  guaranteeing $1.01$ of the asymptotic $\mathrm{BD}$}
    \label{tab:dap_and_bfrac_for_eta}
    \centering
    \def\arraystretch{1.3}
    \begin{tabular}{
      >{\centering\arraybackslash}m{0.15\linewidth}<{}
      |>{\centering\arraybackslash}m{0.15\linewidth}<{}
      |>{\centering\arraybackslash}m{0.15\linewidth}<{}
    |>{\centering\arraybackslash}m{0.15\linewidth}<{}
      |>{\centering\arraybackslash}m{0.15\linewidth}<{}
    }
         \multirow{2}{*}{\shortstack{Array\\geometry}} & \multicolumn{2}{c|}{$\mathrm{SIMO/MISO}$} &  \multicolumn{2}{|c}{$\mathrm{MIMO}$} \\
         \cline{2-5}
         & $D_{\lambda, \min}$ & $\mathrm{B_{f, min}}$ & $D_{\lambda, \min}$ & $\mathrm{B_{f, min}}$ \\
         \hline
         ULA & $41.9$ & $0.045$ & $30$ & $0.063$\\
         \hline
         UCA & $34.6$ & $0.054$ & $25$ & $0.075$ \\
         \hline
         URA & $59.9$ & $0.031$ & $42.6$ & $0.044$ \\
         \hline
         UPCA & $42.7$ & $0.044$ & $30.8$ & $0.061$ \\
    \end{tabular}
\end{table}

\section{Performance improvement metrics}
\label{sec:perf_improv_metrics}
Based on the ambiguity function approximation, this section provides insight into the sensing performance improvements offered by the near-field operation for the considered array geometries. The key performance metrics used in sensing system analysis and design are range resolution, peak-to-sidelobe level (PSL), and integrated sidelobe level (ISL). 

\subsection{Resolution}
The resolution provided by the NF beamfocusing is range-dependent, with the beamdepth function growing very rapidly as the distance from the array increases \eqref{eq:3dB_bd_vs_d}. Generally, the range-dependent resolution is undesirable; therefore, the addition of bandwidth is crucial to provide uniform resolution within a specified region. Depending on the amount of bandwidth, the near-field region might be separated into two subregions in which the resolution is dominated by the NF beamfocusing or bandwidth. As bandwidth is a scarce resource, one should expect NF systems where the NF-dominated resolution region is always present.

\begin{figure}[t]
    \centering
    \includegraphics[width=\linewidth]{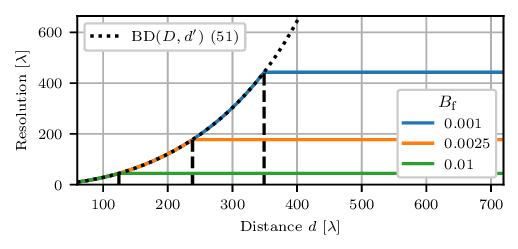}
    \caption{Range resolution in the near-field region for $D_{\lambda} = 50$ and different values of fractional bandwidth $B_{\mathrm{f}}$.}
    \label{fig:res_vs_d_w_bw}
\end{figure}

The boundary distance which separates the near-field and the bandwidth resolution dominated regions can be calculated by equating the BD \eqref{eq:3dB_bd_vs_d} with the resolution due to bandwidth and solving for distance $d$, arriving at
\begin{align}
    d'_{\mathrm{N/B}} &= \sqrt{\frac{1.772 D^4}{4 \alpha B_{\mathrm{f}} D^2 + 0.433 \alpha^2 \lambda^2 }}.
\end{align}
For $ 1.2D < d'_{\mathrm{N/F}} < d_{\mathrm{FA}} / \alpha $ the NF dominated region spans $[1.2D, d'_{\lambda}] $ and the BW dominated region is $[d'_{\mathrm{N/F}}, d_{\mathrm{FA}} / \alpha]$. Based on this division, one can calculate the percentage of the NF region where the resolution is better than the one provided by BW (NF-dominated region). Fig. \ref{fig:res_vs_d_w_bw} presents the resolution in the NF region offered by an NF system with bandwidth. Introducing bandwidth provides uniform sensing resolution for distances at which the resolution due to beamfocusing is insufficient.

\subsection{Peak-to-sidelobe level (PSL)}

\begin{figure}[t]
    \centering
    \includegraphics[width=\linewidth]{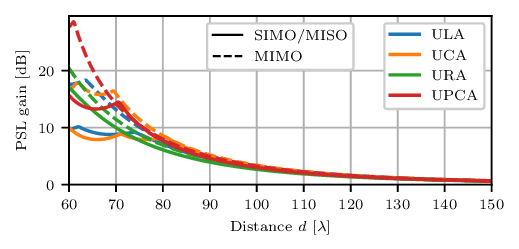}
    \caption{PSL gain as a function of distance for $D_{\lambda} = 50$ and different configurations.}
    \label{fig:psl_gain_vs_d}
\end{figure}

Consider a system with a bandwidth that guarantees identical resolution due to NF beamfocusing and BW. The range-dependent AF modulates the waveform ambiguity function, improving the PSL. However, as the range increases, the width of the NF AF beam that attenuates the sidelobes widens and the PSL improvement diminishes. Fig. \ref{fig:psl_gain_vs_d} illustrates the range-dependent PSL gain provided by the NF.
Notice that the PSL gain is not limited by the PSL of NF AF, since the waveform ambiguity function and NF AF are not correlated, resulting in greater improvement than the measured PSL levels of the NF AF alone. 
The MIMO configuration offers a twofold improvement in the PSL gain in decibel scale due to squaring of the NF AF.

Consider a scenario in which the near-field range-dependent resolution is treated beneficial byproduct, with the key objective of the system to provide uniform PSL performance across all distances. Introducing bandwidth and consequently the waveform ambiguity function helps to suppress the undesirable sidelobe levels of the NF AF.
Since the total ambiguity function is the product of the waveform ambiguity function and the NF AF, the sidelobe performance can be further improved by conventional frequency-domain windowing of the received signal spectrum.
Windowing in the fast-time domain reduces sidelobes at the cost of worsened resolution due to mainlobe broadening. The broader mainlobe of the autocorrelation function due to windowing makes it less sensitive to horizontal shifts, resulting in a relaxed bandwidth-aperture product constraint. 
For OFDM, the windowed waveform ambiguity function is obtained by circular convolution with the inverse Fourier transform of the discrete window
\begin{align}
    \chi_{B,\mathrm{w}} = \chi_{B} \circledast \mathrm{IFFT}(W(k)),
\end{align}
where $W(k)$ is the window function per $k$-th subcarrier.
\begin{table}[t]
    \caption{Window function parameters}
    \label{tab:wdw_func}
    \centering
    \def\arraystretch{1.2}
    \begin{tabular}{
      >{\centering\arraybackslash}m{0.25\linewidth}<{}
      |>{\centering\arraybackslash}m{0.15\linewidth}<{}
      |>{\centering\arraybackslash}m{0.2\linewidth}<{}
    }
         Window & Relative resolution & PSL [dB] \\
         \hline
         Rectangular & 1 & -13.26\\
         \hline
         Hamming & 2 & -43.68  \\
         \hline
         Hann & 2 &  -31.47 \\
         \hline
         Blackman & 3 & -58.11 \\
    \end{tabular}
\end{table}
The following discusses the minimum bandwidth that guarantees PSL level in the near-field, not worse than the typical PSL levels observed when windowing is applied. The considered windows and their parameters are provided in Table \ref{tab:wdw_func}. 
The analysis is performed for asymptotic NF AF, which holds for large arrays that offer a low percentage of the asymptotic resolution. First, the minimum fractional bandwidth is computed $B_{{\mathrm{f}, \min}_\mathrm{PSL} }$ that guarantees that the PSL levels observed in the NF span are better or equal to the ones typical for the far-field. This means that the PSL levels of the total NF ambiguity function are lower than or equal to the far-field PSL values specified in Tab. \ref{tab:wdw_func}. 
Next, the obtained $B_{{\mathrm{f}, \min}_\mathrm{PSL} }$ is compared with \eqref{eq:bf_min}, the minimum fractional bandwidth that guarantees the required PSL with only bandwidth $B_{\mathrm{f}, \min}$. The resolution loss due to windowing is taken into account when calculating $B_{\mathrm{f}, \min}$. The ratio $B_{\mathrm{f}, \min}/ B_{{\mathrm{f}, \min}_\mathrm{PSL} }$ is computed as a metric indicating how much less bandwidth is required in the near-field to provide the far-field PSL performance as compared to the system where resolution is solely determined by bandwidth. 
\begin{figure}[t]
    \centering
    \includegraphics[width=\linewidth]{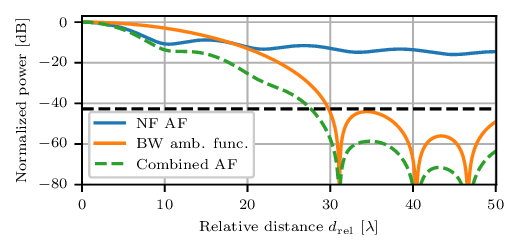}
    \caption{Ambiguity function of a ULA SIMO/MISO NF system with bandwidth equal to $B_{{\mathrm{f}, \min}_\mathrm{PSL}}$, Hamming window and target at minimum distance from the array. The dashed black line denotes the PSL of a Hamming window.}
    \label{fig:min_bf_af}
\end{figure}

\begin{table}[b]
    \caption{$B_{\mathrm{f}, \min}/ B_{{\mathrm{f}, \min}_\mathrm{PSL} }$ ratio for SIMO/MISO}
    \label{tab:bfrac_ratio_per_psl_simo_miso}
    \centering
    \def\arraystretch{1.2}
    \begin{tabular}{
      >{\centering\arraybackslash}m{0.15\linewidth}<{}
        |>{\centering\arraybackslash}m{0.15\linewidth}<{}
        |>{\centering\arraybackslash}m{0.15\linewidth}<{}
        |>{\centering\arraybackslash}m{0.15\linewidth}<{}
        |>{\centering\arraybackslash}m{0.15\linewidth}<{}
    }
         \shortstack{Array\\geometry}
         & Rectangular & Hamming & Hann & Blackman \\
         \hline
         ULA & $2.33$ & $2.76$ & $2.51$ & $3.25$ \\
         \hline
         UCA & $2.51$ & $1.27$ & $1.42$ & $1.22$ \\
         \hline
         URA & $\infty$ & $2.84$ & $2.61$ & $3.32$ \\
         \hline
         UPCA & $\infty$ & $1.39$ & $1.6$ & $1.32$ \\
    \end{tabular}
\end{table}
\begin{table}[b!]
    \caption{$B_{\mathrm{f}, \min}/ B_{{\mathrm{f}, \min}_\mathrm{PSL} }$ ratio for MIMO}
    \label{tab:bfrac_ratio_per_psl_mimo}
    \centering
    \def\arraystretch{1.2}
    \begin{tabular}{
      >{\centering\arraybackslash}m{0.15\linewidth}<{}
        |>{\centering\arraybackslash}m{0.15\linewidth}<{}
        |>{\centering\arraybackslash}m{0.15\linewidth}<{}
        |>{\centering\arraybackslash}m{0.15\linewidth}<{}
        |>{\centering\arraybackslash}m{0.15\linewidth}<{}
    }
         \shortstack{Array\\geometry}
         & Rectangular & Hamming & Hann & Blackman \\
         \hline
         ULA & $\infty$ & $2.84$ & $2.61$ & $3.32$ \\
         \hline
         UCA & $\infty$ & $2.15$ & $2.56$ & $1.99$ \\
         \hline
         URA & $\infty$ & $5.05$ & $\infty$ & $3.51$ \\
         \hline
         UPCA & $\infty$ & $2.68$ & $4.29$ & $2.28$ \\
    \end{tabular}
\end{table}

Table \ref{tab:bfrac_ratio_per_psl_simo_miso} and \ref{tab:bfrac_ratio_per_psl_mimo} present the bandwidth ratio for SIMO/MISO and MIMO configurations, respectively. The $\infty$ symbol denotes that the PSL offered by the NF AF already satisfies the target far-field PSL and the required bandwidth $B_{{\mathrm{f}, \min}_\mathrm{PSL} }$ is zero. The ratio varies across antenna geometries due to the different AF functions that determine the shape and periodicity of the minima and the nulls of the AF. The MIMO processing reduces the required bandwidth due to the steeper AF, with some exceptions, notably ULA and URA. For ULA and URA, the gains are limited due to the lack of nulls in the NF AF, as can be seen in Fig. \ref{fig:af_per_geom}. The squaring of the NF AF due to MIMO processing increases the depth of the minima, which need to be compensated (flattened out)  with the waveform ambiguity function, so as not to create a sidelobe. Fig. \ref{fig:min_bf_af} illustrates the SIMO/MISO ULA, the ambiguity function due to NF, BW and the composite AF for $B_{{\mathrm{f}, \min}_\mathrm{PSL}}$.
To achieve the same PSL performance as in the far-field, the required bandwidth in the near-field is only a fraction of the bandwidth associated with the resolution provided by NF beamfocusing.

\subsection{Integrated sidelobe level (ISL)}

\begin{figure}[h]
    \centering
    \includegraphics[width=\linewidth]{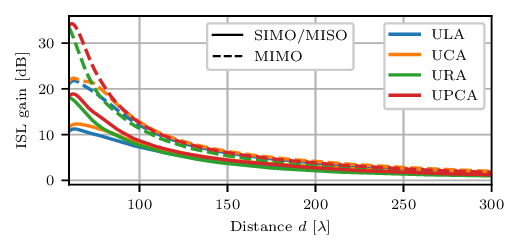}
    \caption{ISL gain as a function of distance for $D_{\lambda} = 50$ and different configurations.}
    \label{fig:isl_gain_vs_d}
\end{figure}

The NF beamfocusing also improves the integrated sidelobe levels by suppressing higher-order sidelobes of the ambiguity function due to waveform. Consider a system where the resolution due to bandwidth is equal to the minimum beamdepth (beamfocusing resolution). The range-dependent NF AF effectively suppresses the sidelobes in the close proximity to the array. However, as the distance increases, the ISL gains diminish due to the widening beamdepth of the NF AF function. Fig. \ref{fig:isl_gain_vs_d} illustrates the ISL gain as a function of distance. The MIMO configuration offers approximately a twofold improvement in the gain in decibel scale due to squaring of the NF AF.

\section{Conclusion}

This paper presents an approximation of the near-field ambiguity function by separating it into a product of the ambiguity function due to the waveform and the near-field array factor. The approximation constraint is derived as a bandwidth-aperture product for different configurations, SIMO/MISO and MIMO and array geometries. Separation of the ambiguity function into two factors allows for analyzing the benefits that the near-field offers to systems with bandwidth and vice versa. When the resolution due to bandwidth is comparable to the near-field beamfocusing, the PSL and ISL metrics of the composite AF are greatly improved as the NF AF serves as a mask, suppressing the sidelobes. The gains are observed to diminish with growing distance due to the widening of the beamdepth function. Alternatively, when considering NF-centric systems, the bandwidth can be used to improve the poor sidelobe levels of the NF AF. The minimum bandwidth guaranteeing uniform sidelobe performance matching the far-field is shown to be lower than the one corresponding to the beamfocusing resolution.

\bibliographystyle{IEEEtran}
\bibliography{IEEEabrv.bib, biblio.bib}

\end{document}